\documentclass[Afour,sagev,times]{sagej}
\usepackage[switch,columnwise]{lineno}
\usepackage[dvipsnames,svgnames,table]{xcolor}
\usepackage{sectsty}
\usepackage{multirow}
\usepackage{authblk}
\usepackage{relsize}
\usepackage[T1]{fontenc} 
\usepackage{fourier} 
\usepackage[english]{babel} 
\usepackage{amsmath,amsfonts,amsthm, amssymb} 
\usepackage{graphicx,placeins,subcaption}
\usepackage{wrapfig}
\usepackage{bm}
\usepackage{soul} 
\usepackage{lipsum} 
\usepackage{sectsty} 
\usepackage{fancyhdr} 
\usepackage[T1]{fontenc} 
\usepackage[english]{babel} 
\usepackage{framed}
\usepackage{bm}
\usepackage{fancyhdr}
\usepackage{etoolbox}
\usepackage{titlesec}
\usepackage{titletoc}
\usepackage{float}
\usepackage{xcolor}
\usepackage{moreverb,url}

\usepackage[colorlinks,bookmarksopen,bookmarksnumbered,citecolor=red,urlcolor=red]{hyperref}

\newcommand\BibTeX{{\rmfamily B\kern-.05em \textsc{i\kern-.025em b}\kern-.08em
T\kern-.1667em\lower.7ex\hbox{E}\kern-.125emX}}

\def\volumeyear{2016}

\begin{document}

\runninghead{}

\title{Ultrasonic monitoring of stress and cracks of the 1/3 scale mock-up of nuclear reactor concrete containment structure}

\author{Qi Xue\affilnum{1}, Eric Larose\affilnum{1}, Ludovic Moreau\affilnum{1}, Romain Thery\affilnum{1,2}, Odile Abraham\affilnum{2}, Jean-Marie Henault\affilnum{3}}

\affiliation{\affilnum{1}Univ. Grenoble Alpes, CNRS, ISTerre, CS 40700, 38058 Grenoble, France\\
\affilnum{2}Uni. Eiffel, GERS-GeoEND, all\'ee des ponts et chauss\'ees - CS5004, 44344 Bouguenais Cedex, FRANCE \\
\affilnum{3} EDF R\&D, Chatou, France}

\corrauth{\'Eric Larose, ISTerre,
Grenoble, France.}

\email{eric.larose@univ-grenoble-alpes.fr}

\begin{abstract}
To evaluate the stress level and damage of a reinforced  concrete containment wall 
(similar to those used in nuclear power plants) and its reaction to pressure variations,
we implemented successive ultrasonic experiments on the exterior surface of the containment wall in the gusset area 
for three consecutive years (2017, 2018 and 2019).
During each experiment, the pressure inside the containment wall increased gradually from 0 MPa to 0.43 MPa and then decreased back to 0 Mpa.
From the analysis of the ultrasonic coda waves obtained in the multiple scattering regime (80-220 kHz), we performed Coda Wave Interferometry to calculate the apparent velocity changes in the structure (denoted by $dV/V_a$) and Coda Wave Decorrelation (DC) measurements 
to produce 3D cartographies of stress and crack distribution.
From three source-receiver pairs, located at the top, middle and bottom of the experimental region,
we observe that coda waves dilate, shrink and remain almost unchanged, respectively. 
This corresponds to the decreasing, increasing and invariant pressure inside the concrete.
The comparison of three years' results demonstrates that the variation of $dV/V_a$ and DC 
under the same pressure test increases through the years, which indicates the progressive deterioration and aging of the concrete.
From a large collection of source-receiver pairs at different times, the spatial-temporal variations of $dV/V_a$ and DC are then used to produce a map of the structural velocity and scattering changes, respectively.
We observe a decreasing velocity on the top part and an increasing in the middle one, which is in line with the $dV/V_a$ analysis.
The reconstructed scattering changes (or structural changes) highlight the active region during the inflation-deflation procedure, corresponding to the opening and closing (and sometimes the development) of cracks.
The larger magnitude in 2019 than in 2017 indicates the increasing damage in the concrete.
\end{abstract}


\maketitle

\pagestyle{fancy}

\section{Introduction}
Concrete containments are generally used to protect the nuclear reactor of a nuclear power plant 
in case of any leakage of the nuclear fuel or waste, or from possible external aggression.
The humidity, temperature and/or pressure changes induce the deformation of the concrete shell and, together with additional chemical processes, progressively alter the mechanical integrity of the structure.
Small cracks (millimeter scale) are generated and progressively develop during  the lifetime of the structure.
The assessment of the state of damage and the characterization of cracks are vital issues
for the security of the nuclear power plant in case of emergency, and for the estimation of the residual life time.
In France, a leak test of each concrete containment is enforced by the law every ten years\footnote{https://www.legifrance.gouv.fr/jorf/id/JORFTEXT000000469544},
during which the pressure inside the containment is increased from 0 MPa to about 0.43 MPa
and then decreased back to 0 MPa.

In order to study the aging of the concrete wall and follow the evolution of the leak rate owning to drying creep and loss
of pre-stress in a realistic environment, \'{E}lectricit\'{e} de France (EDF) launched a research program on a 1/3 scale mock-up, called VeRCoRs \cite{masson2013objectives,mathieu2018temperature,henault2018characterize}.
This  mock-up is a double walled concrete containment building, as a 1:3 scale specimen representative for
24 (about 40 \%) of the French reactors.  The present work explores the applicability and performance of a
method based on diffuse ultrasound analysis to image and quantify stress changes and cracks within the structure at VeRCoRs during a standard internal pressure test.

To this end, we take advantage of the high sensitivity of multiply scattered waves at the ultrasonic scale, for frequencies of few hundreds of kHz.
The high heterogeneity of the concrete results in the loss of ballistic and coherent properties 
of the waveforms after a propagation over a few wavelengths.
The late-arriving ultrasonic signals, also named coda waves, 
are known to be highly repeatable and sensitive to small perturbations of the medium. 
At the seismic scale, the sensitivity of the coda wave has been extensively exploited to detect small changes in the Earth's crust, 
e.g., fault zones \cite{wegler2007fault, brenguier2008postseismic} 
or volcanic areas \cite{gret2005monitoring, sens2006passive, brenguier2008towards}.

Coda wave interferometry (CWI), as introduced in \cite{poupinet1984monitoring} and 
further developed in \cite{snieder2002coda,snieder2006theory}, is a technique to detect slight velocity perturbations 
in a highly heterogeneous medium, by comparing coda waves recorded before and after a perturbation. 
A quantitative way to compare two waveforms  is to compute their correlation coefficient (CC).
One way of applying CWI consists in stretching or shrinking the waveform to match two coda signals, 
i.e., to maximize the CC in a given time window, with a dilation coefficient.
This coefficient, associated to the local velocity change \cite{lobkis2003coda}, 
is generally referred to as the apparent relative velocity change $dV/V_a$. 
Alternatively, it can be useful to calculate the decorrelation coefficient (DC) instead of the CC, 
because the DC can be linked to small structural changes \cite{planes2015imaging}.
The $dV/V_a$ and DC variables are now widely used to detect ultrasonic velocity changes in a concrete structure
induced by thermal or stress variations \cite{larose2006observation,larose2009monitoring,schurr2011damage},
and/or to detect micrometer scale cracks \cite{hilloulin2014small}. The reader can refer to \cite{planes2013review} for a review.

The relation between the $dV/V_a$ and the local velocity change is proposed in \cite{pacheco2005time} by comparing wave propagation with random walk.
The relation between DC and the local scattering change (for example, induced by cracks) follows a similar formula proposed later \cite{larose2010locating}.
The reader may refer to \cite{planes2014decorrelation} for a detailed explanation and proof.
These two formulas have the same form: an integral equation with the same sensitivity kernel.
The sensitivity kernel is a function of the source, receiver, space and time, that statistically estimates the density of time spent by the diffuse wave at a given region of the medium. It can be estimated either from i) the explicit solution of diffusive or radiative-transfer approximation of ultrasonic waves for simple geometries (e.g., infinite plate or cuboid) \cite{planes2014decorrelation}, ii)
numerical methods (e.g., finite difference or finite element) for any shape \cite{xue2019locating}
or iii) directly solving the wave equation \cite{kanu2015numerical}.
By combining information from several sources and receivers, we are able to obtain thousands of $dV/V_a$ and DC values
between different source-receiver pairs in different times, which can be used to image 
the velocity and scattering changes through solving the integral equation.
Please refer to \cite{planes2015imaging} for a numerical simulation, \cite{zhang2016diffuse} and \cite{zhan2020three}
for laboratory experiments and \cite{zhang2018three} for a field experiment.

We implemented similar diffuse ultrasonic experiments on the outer surface of the gusset area of VeRCoRs (see Fig.~\ref{fig:config}(a))
during the internal pressure test over three consecutive years, in order to detect cracks and the effect of damage on stress changes during each test. From the analysis of the $dV/V_a$ and DC for three different source-receiver pairs between the initial state and other states during the pressure test, we aim at observing that different parts of the wall are expected to have different responses to the same inflation-deflation procedure. We also aim at quantifying the evolution of $dV/V_a$ and DC versus aging and damage over the years.
An imaging inversion procedure is then applied to obtain 3D cartographies of relative velocity changes and density of probability of cracks.

The comparison of different years' results may quantify the deterioration of the concrete:
the sensitivity of $dV/V_a$ to internal stress is expected to increase with time (thus with damage)
as observed elsewhere \cite{planes2013review} and the decorrelation is expected to show some non-reversible (thus cumulative) behavior.

\section{Experimental setup}

\begin{figure*}[!htb]
\centering
\begin{subfigure}[p]{.265\textwidth}
    \centering
   	\includegraphics[width=\textwidth]{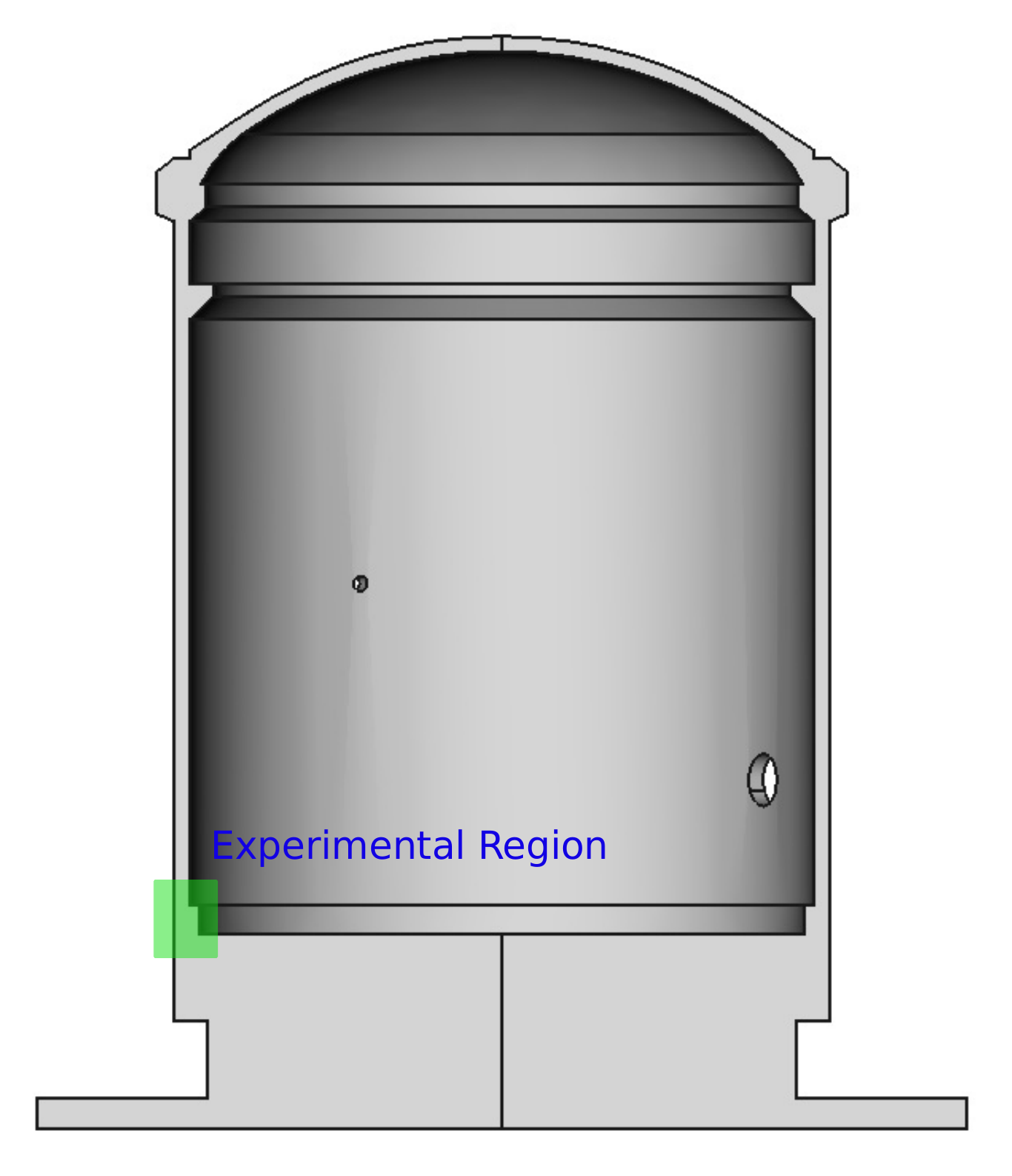}
	\captionsetup{justification=centering}
   	\caption{}
\end{subfigure}~
\begin{subfigure}[p]{.35\textwidth}
    \centering
   	\includegraphics[width=\textwidth]{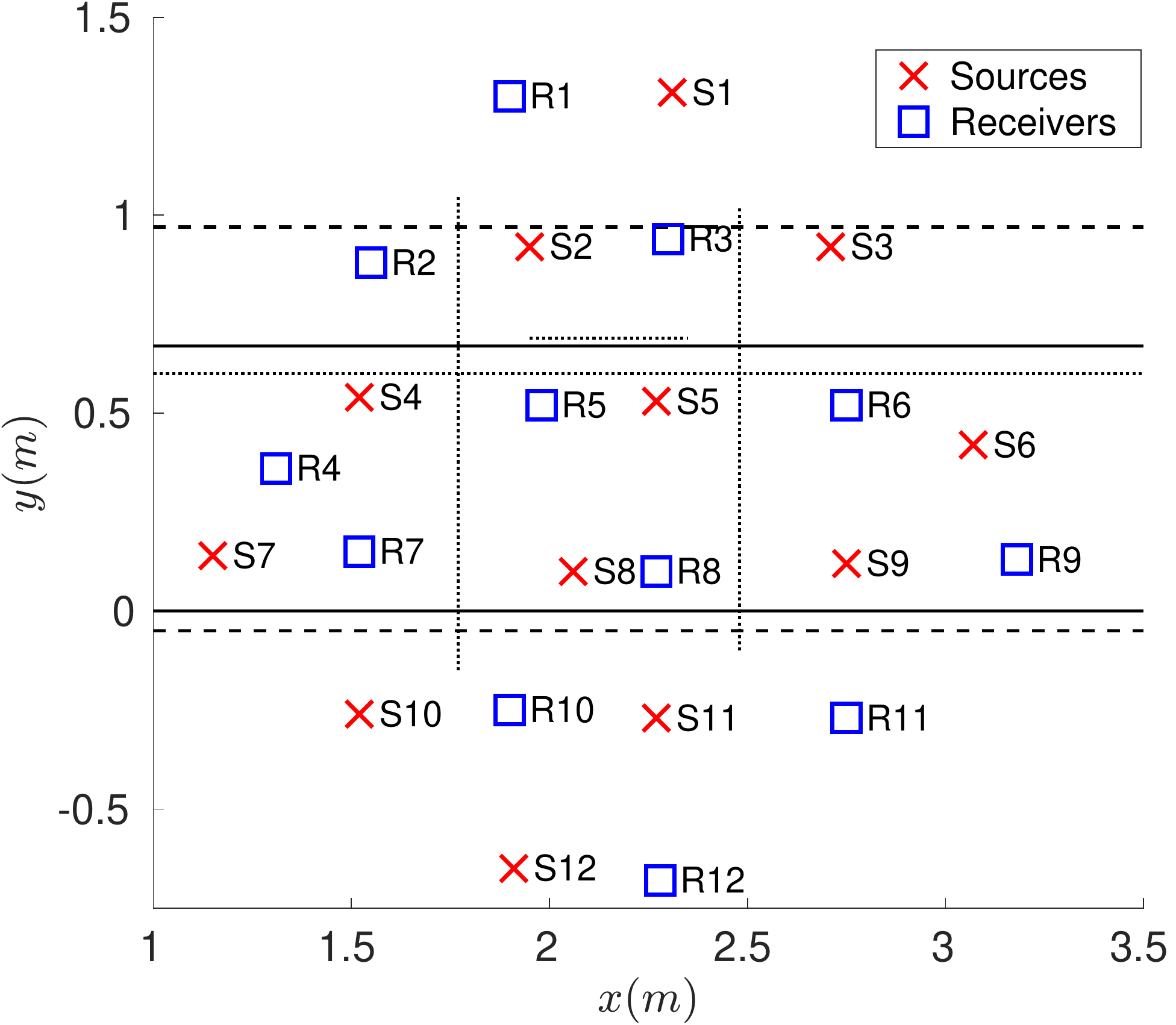}
	\captionsetup{justification=centering}
   	\caption{}
\end{subfigure}~
\begin{subfigure}[p]{.35\textwidth}
    \centering
   	\includegraphics[width=\textwidth]{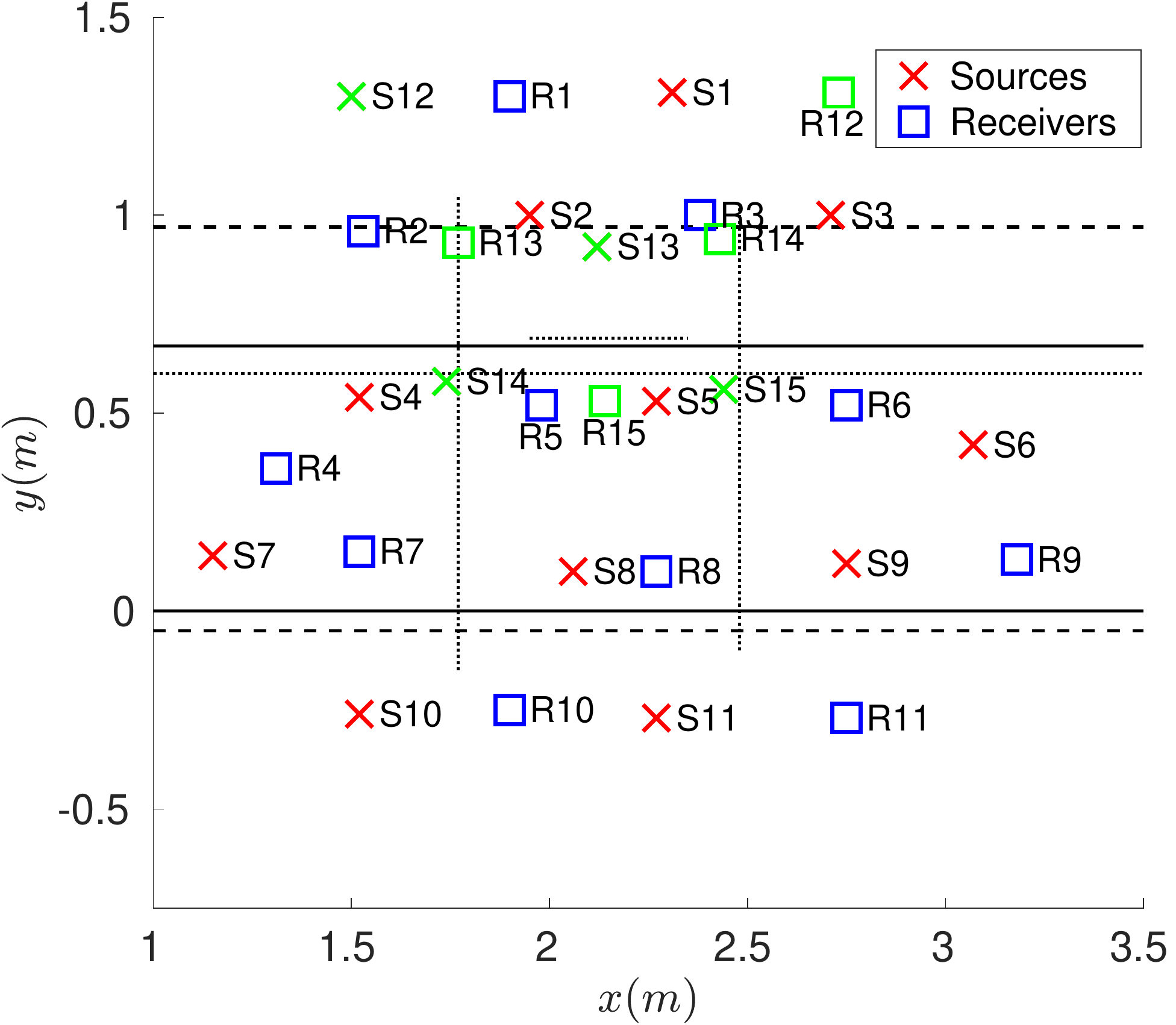}
   	\captionsetup{justification=centering}
   	\caption{}
\end{subfigure}
\caption{Configuration of the experiment. (a) Sketch of the concrete structure and the experimental region (green). We focus on the lower corner (the gusset) of the exterior surface of VeRCoRs, a 1/3 replica of a nuclear containment wall.
(b) source and receiver positions in 2017 and 2018. (c) source and receiver positions in 2019 with new transducer positions in green.
The horizontal solid lines represent the change of thickness of the wall. 
The thickness of the upper part is 40 cm and the middle part (gusset) 60 cm. The bottom layer is the base slab.
The dashed horizontal lines are positions of the construction joint.
Dotted lines are positions of cracks on the surface of the wall observed in 2018.}
\label{fig:config}
\end{figure*}

VeRCoRs is a 1/3 scale experimental mock-up of a nuclear reactor containment 
(see Fig.~\ref{fig:config}(a)) built by EDF near Paris\footnote{https://fr.xing-events.com/EDF-vercors-project.html?page=1426249}.
Construction of the mock-up began in 2013 and was completed in March 2016.
The construction procedure is layer by layer.
The horizontal dashed lines in Fig.~\ref{fig:config}(b) and (c) indicate positions of the joint in our experimental region. 
After the construction, two auxiliary air-conditioning units started to impose relative realistic humidity and temperature 
inside and outside the containment to simulate regular operations of a nuclear reactor.
The 1/3 scale was chosen to accelerate aging, ten times faster than in a realistic environment hopefully.

To monitor the properties of the concrete, the pressure inside the structure increases gradually from 0 MPa to 0.43 MPa,
which is followed by a static period, during which the pressure decreases a little bit due to small leakage from cracks, 
and then the whole structure is deflated to its initial state (see  Fig.~\ref{fig:pressure_stretch_DC}(a)). 
This test was repeated several times, twice in 2017, and once in 2018 and 2019,
which corresponds to the standard leak test for real nuclear power plant containment every ten years.
We foresee a large change of the concrete in the first test in 2017,
because it was the first pressure test and a lot of micro and macro-cracks were created.
Therefore, in this paper we only present the second test in 2017,
to take into account the so-called Kaiser effect that has been observed and commented previously in concrete \cite{zhang2012study}.
Modifications observed in the following are thus not related to the fact that the pressure level exceeds the first one.
Our experiment was implemented on the outer surface of the gusset region (a surface of about 2m$\times$2m).
Fig.~\ref{fig:config}(a) provides a sketch of the structure and the experimental region.

Several transducers are glued directly to the concrete to generate and receive ultrasonic signals.
Fig.~\ref{fig:config}(b) shows the positions of sources and receivers in 2017 and 2018, and 
Fig.~\ref{fig:config}(c) shows positions in 2019.
Since the base of the concrete structure (base slab) is extremely thick, 
the signals recorded by the bottom source-receiver pair (S12-R12) remain almost unchanged during the experiment in 2017 and 2018.
We decided to remove them and to add several new transducers to improve the resolution of the upper part in 2019. 
In addition, we did not receive reasonable signals from R1 in 2018 which may be caused by the failure or detachment of the transducer.

During the experiment, a chirp signal with frequency ranging from 80 to 220 kHz is generated, amplified and 
sent into the concrete from each source one after another, and then recorded by all the receivers
in order to reconstruct all the source-receiver impulse responses, through deconvolution of recorded waveforms with the chirp source signal. 
The whole acquisition process takes about 3 minutes, within which the change of concrete can be neglected.
The same process is repeated every hour to obtain measurements at different loading states.

\section{Apparent velocity change and decorrelation analysis}
\label{sec:stretch_dc_analysis}

\begin{figure*}[!htb]
\centering
\includegraphics[width=\textwidth]{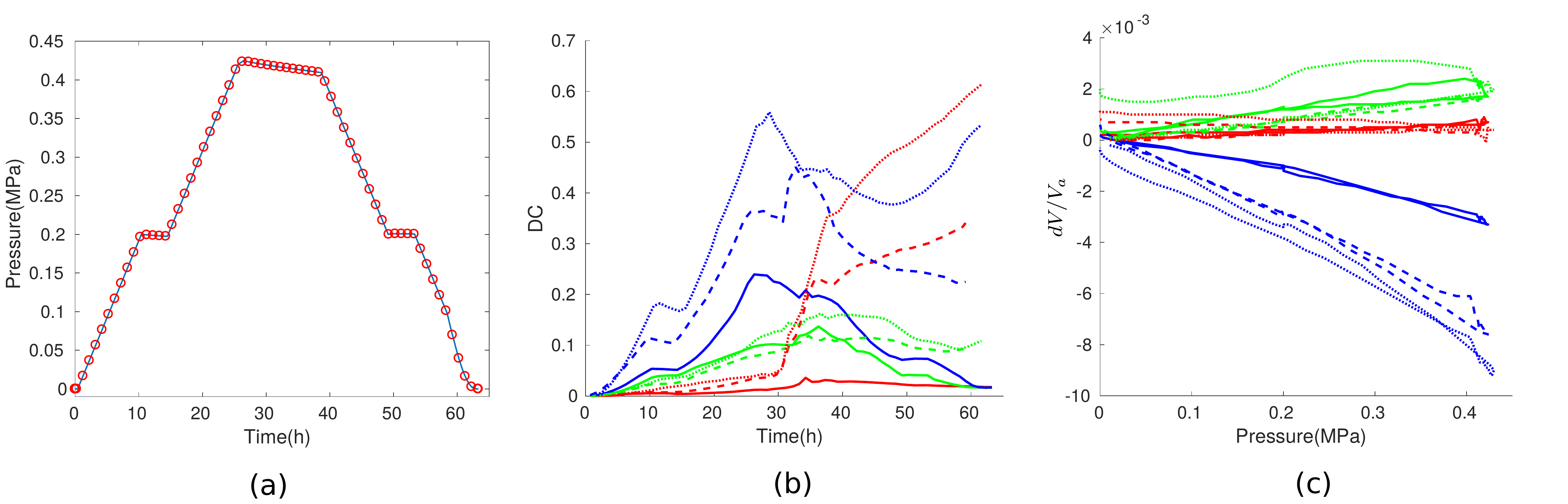}
\caption{(a) pressure variation during the inflation-deflation procedure.
We took the measurement every an hour, during which the pressure has a change of + or - 0.02 MPa 
except in steady states.
(b) DC between the original state and other states;
(c) $dV/V_a$ variation with respect to pressure.
Blue, green and red lines in (b) and (c) correspond to the top, middle and bottom source-receiver pairs 
(S2-R3, S6-R6 and S10-R10 in Fig.~\ref{fig:config}(b)), respectively.
Solid, dashed and dotted lines are results from 2017, 2018 and 2019, respectively.}
\label{fig:pressure_stretch_DC}
\end{figure*}

Let us consider ultrasounds generated and recorded by the same source-receiver pair in two different states: $\phi_1(t)$ and $\phi_2(t)$.
The apparent velocity change $dV/V_a$ on the time interval $[t_1,t_2]$ is defined by
$$dV/V_a = \arg\max_{\alpha}\frac{\int_{t_1}^{t_2}\phi_1((1+\alpha)t)\phi_2(t)dt}
	{\sqrt{\int_{t_1}^{t_2}\phi_1^2((1+\alpha)t)dt\int_{t_1}^{t_2}\phi_2^2(t)dt}},$$
i.e., the amount of stretching that maximizes the cross-correlation (CC) between the signals,
and the DC is one minus CC, i.e.,
$$\text{DC} = 1-\max_{\alpha}\frac{\int_{t_1}^{t_2}\phi_1((1+\alpha)t)\phi_2(t)dt}
	{\sqrt{\int_{t_1}^{t_2}\phi_1^2((1+\alpha)t)dt\int_{t_1}^{t_2}\phi_2^2(t)dt}}.$$

As an example, if the velocity decreases in the second state and if we focus on a peak  at $t_0$ in the wave field $\phi_1$, then the same peak should arrive after $t_0$ in $\phi_2$. Therefore, to match the peak in these two waveforms, $\alpha$ should be negative.
That is to say, a dilation of the time axis (a negative $dV/V_a$) indicates a decreasing  velocity.

To quantitatively study the response of  concrete to pressure changes in different regions and
to compare the state of the concrete over different years, we compute $dV/V_a$ and DC for three source-receiver pairs 
between the original state and other states during the inflation-deflation procedure:
the top pair (S2-R3 in Fig.~\ref{fig:config}), the middle pair (S6-R6) and the bottom pair (S10-R10).

Fig.~\ref{fig:pressure_stretch_DC}(c) shows the value of $dV/V_a$ for source-receiver pairs located at the top (in blue), middle (in green) and bottom (in red) zones of the structure, during inflation cycles from  2017 to 2019. When the the structure is inflated, stress builds up in concrete, as shown in Fig. \ref{fig:7a} and \ref{fig:7b} in appendix B, where we plot tangential stress distribution calculated with linear elastic simulations.  The stress distribution shows traction in the upper part of the structure. This is consistent with what is observed in Fig.~\ref{fig:pressure_stretch_DC}(c), which exhibits a decrease of velocity for the source-receiver pair located at the top region. However, the apparent velocity change remains small in the middle and marginal at the bottom, where the simulation shows a constant stress or compression. 

We have also checked the uncertainties of the apparent velocity changes. Appendix A describes how the uncertainties are calculated, following the formula introduced by \cite{weaver2011precision}. These uncertainties are shown as colored areas in Fig.~\ref{fig:dv_precision}, for the source-receiver pair located at the upper region of the structure. One may see that they remain small despite strong apparent velocity changes.

From 2017 to 2019 one notices an increase in the slope of $dV/V_a$, which suggests that the sensitivity of concrete to the same pressure change
becomes larger. That is to say, the deterioration of the concrete is more pronounced \cite{planes2013review}.
Another interesting phenomenon is that in 2017 the line during inflation almost overlaps with the line during deflation
which means that expansion and contraction of the concrete are fully reversible, because the concrete properties have not evolved significantly between the first ever pressure test in 2017 and the one performed soon after presented here, which is no more the case in 2019.
This might be a good criterion to quantify the damaging impact of the pressure
tests, and to test the quality of the concrete.
Similar conclusions can be drawn for the middle and bottom pairs, though a much lower resolution: 
the middle region experiences a contraction and the pressure in the bottom region keeps almost unchanged.
We leave the precision study of the $dV/V_a$ due to changes in wave speed in the Appendix.

In 2017, the concrete retrieves its initial state after the whole inflation-deflation procedure, 
as indicated by DC (solid lines) in Fig.~\ref{fig:pressure_stretch_DC}(b), which is not the case in 2018 and 2019.
We conclude from lines with the same color that the change of DC becomes more irreversible as time goes by,
which is caused by the cumulative deterioration of the concrete (including micro-cracks and cracks).
The large value of DC between the initial and final states implies that the damage induced by pressure changes is permanent: some cracks open and close reversibly while others develop further and the crack surface is altered such that they do not close exactly back to their initial state.
What's also interesting is the bottom pair: we do not observe significant pressure changes in the bottom region from $dV/V_a$, but they do suffer dramatic scattering (or structural) changes (see Fig.~\ref{fig:pressure_stretch_DC}(b)).

\section{3D cartography of velocity and structural changes}
\label{sec:inverse_dv_cs}

\begin{figure*}[bht]
\centering
\begin{subfigure}[p]{.33\textwidth}
    \centering
   	\includegraphics[width=\textwidth]{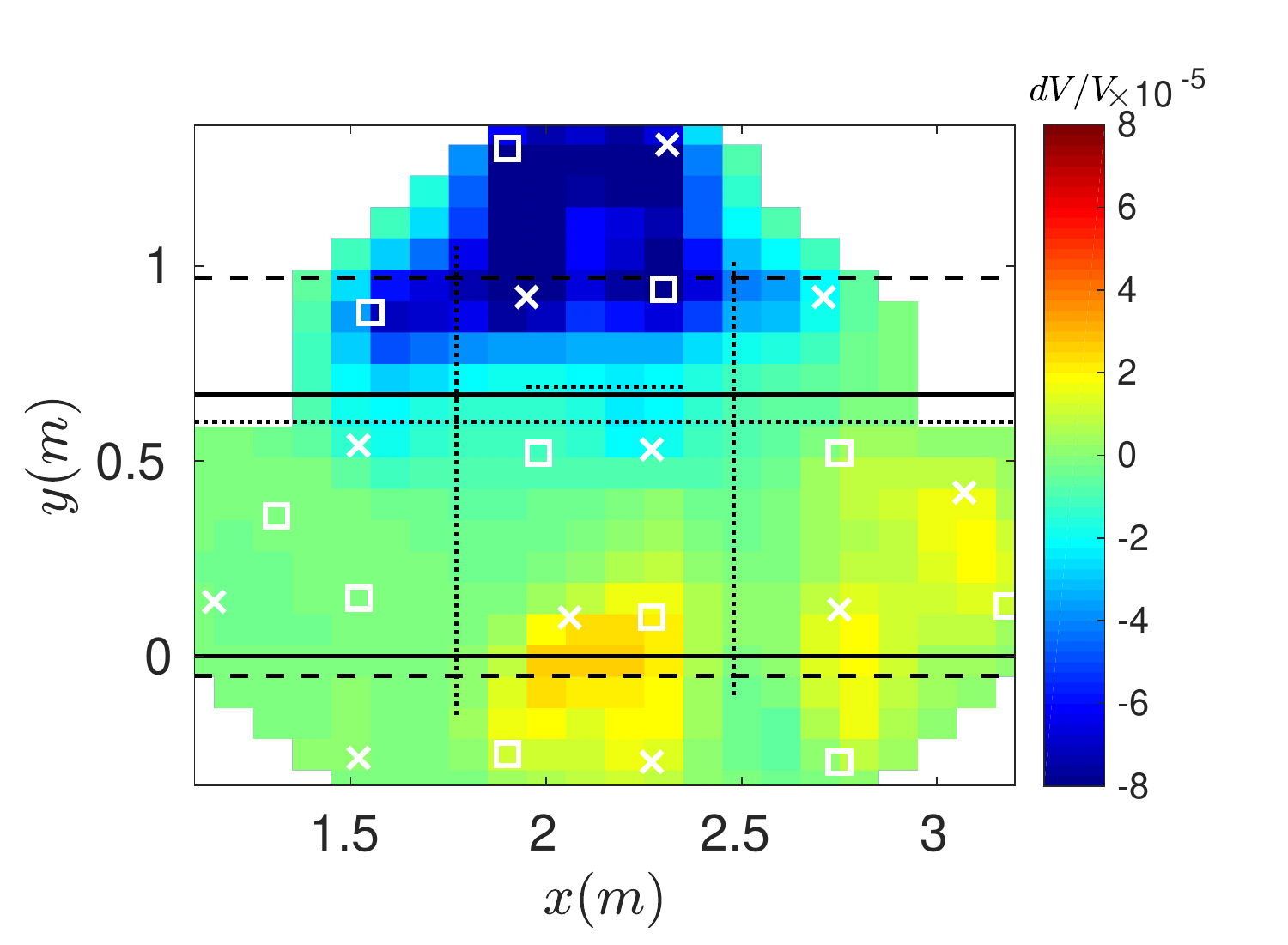}
	\captionsetup{justification=centering}
   	\caption{}
\end{subfigure}~
\begin{subfigure}[p]{.33\textwidth}
    \centering
   	\includegraphics[width=\textwidth]{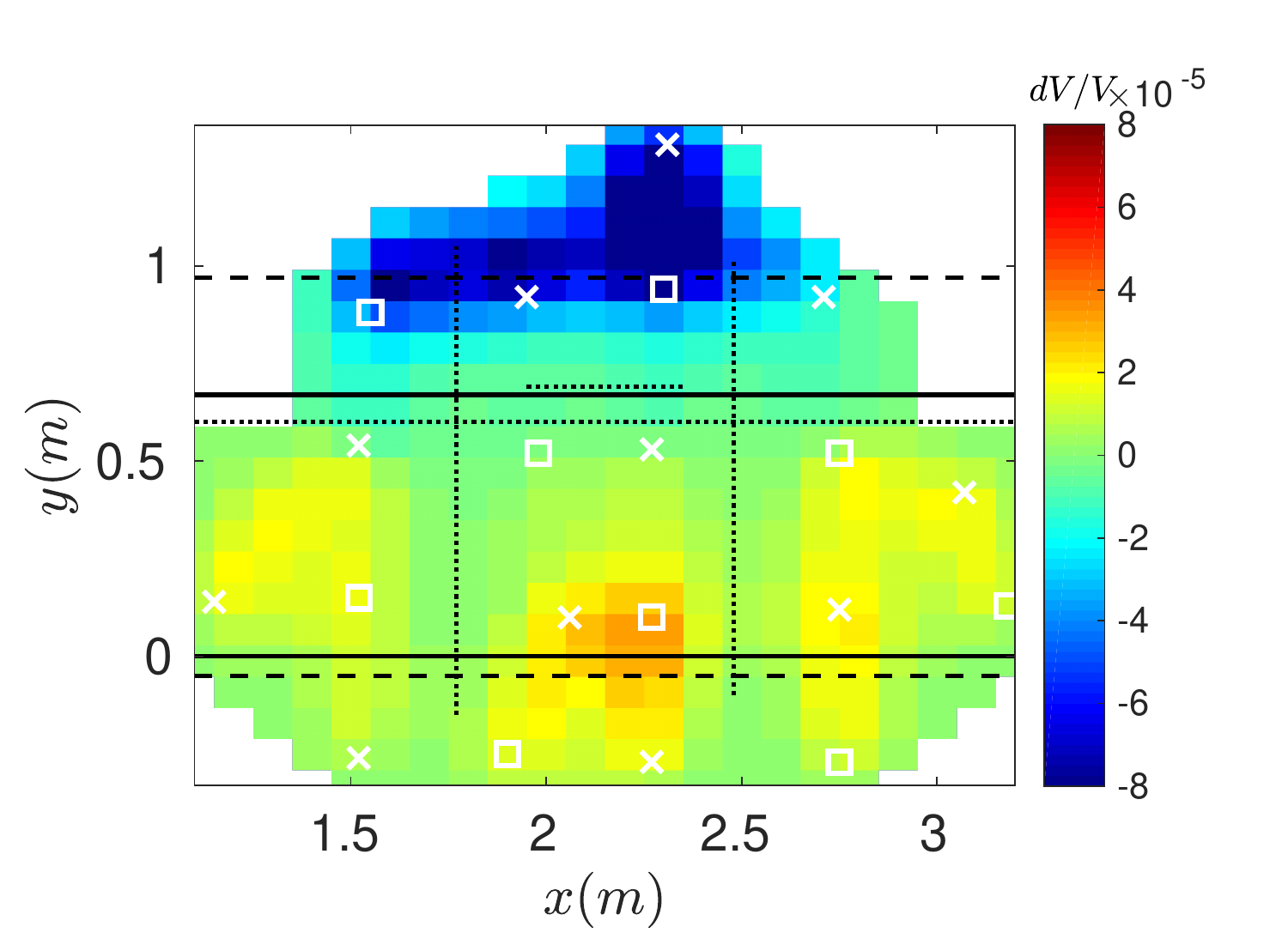}
   	\captionsetup{justification=centering}
   	\caption{}
\end{subfigure}~
\begin{subfigure}[p]{.33\textwidth}
    \centering
   	\includegraphics[width=\textwidth]{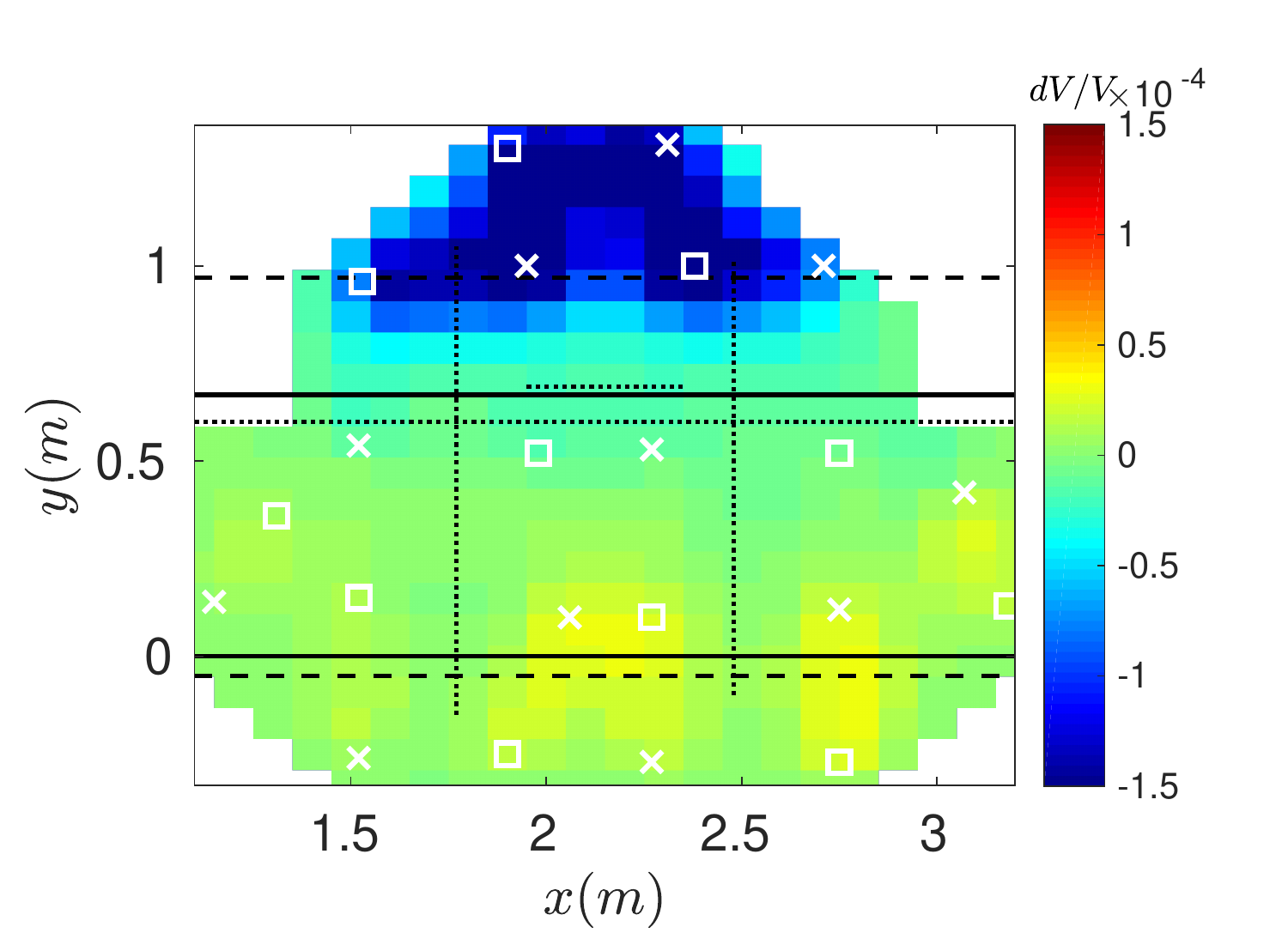}
   	\captionsetup{justification=centering}
   	\caption{}
\end{subfigure}
\begin{subfigure}[p]{.33\textwidth}
    \centering
   	\includegraphics[width=\textwidth]{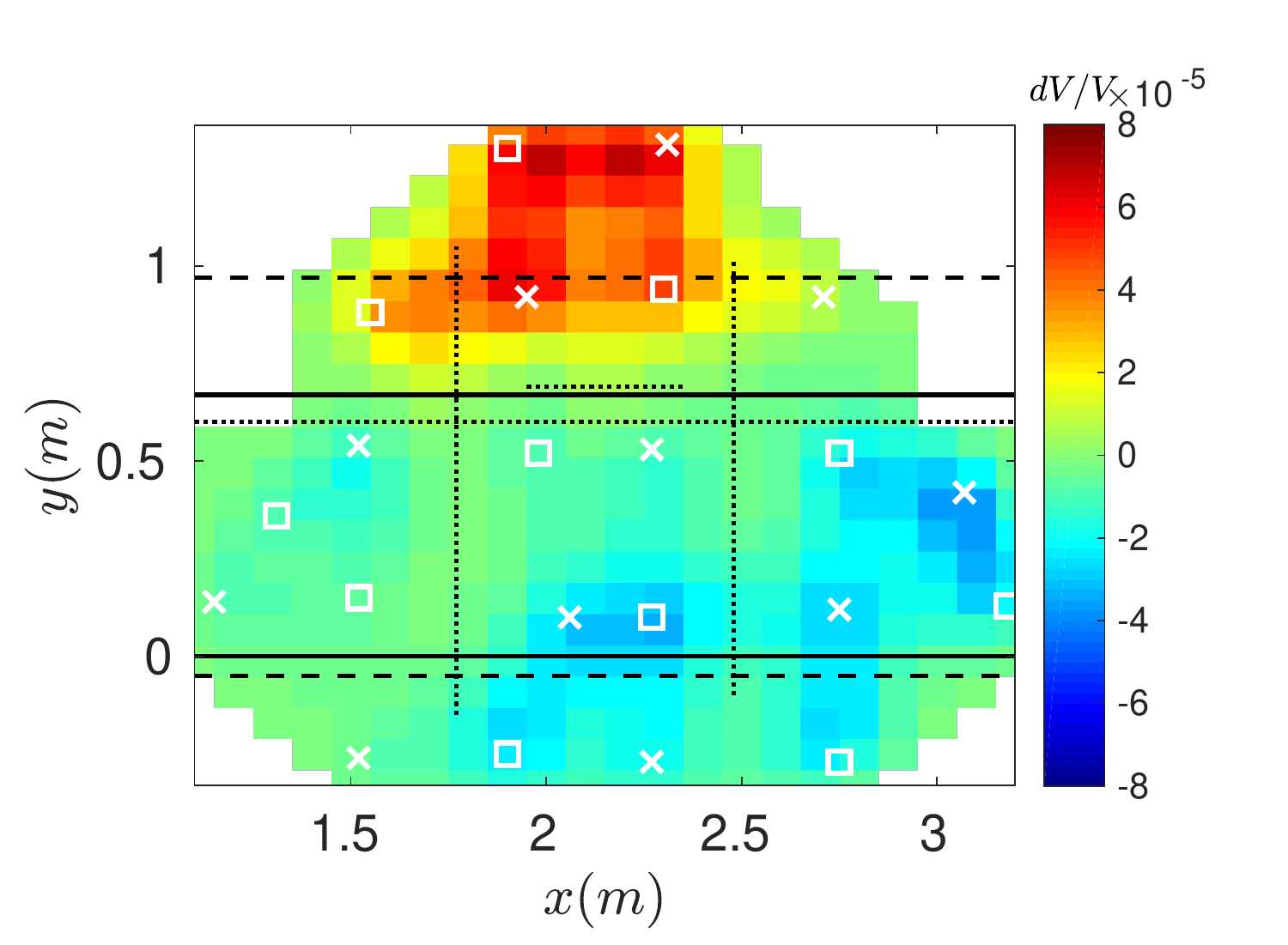}
   	\captionsetup{justification=centering}
   	\caption{}
\end{subfigure}~
\begin{subfigure}[p]{.33\textwidth}
    \centering
   	\includegraphics[width=\textwidth]{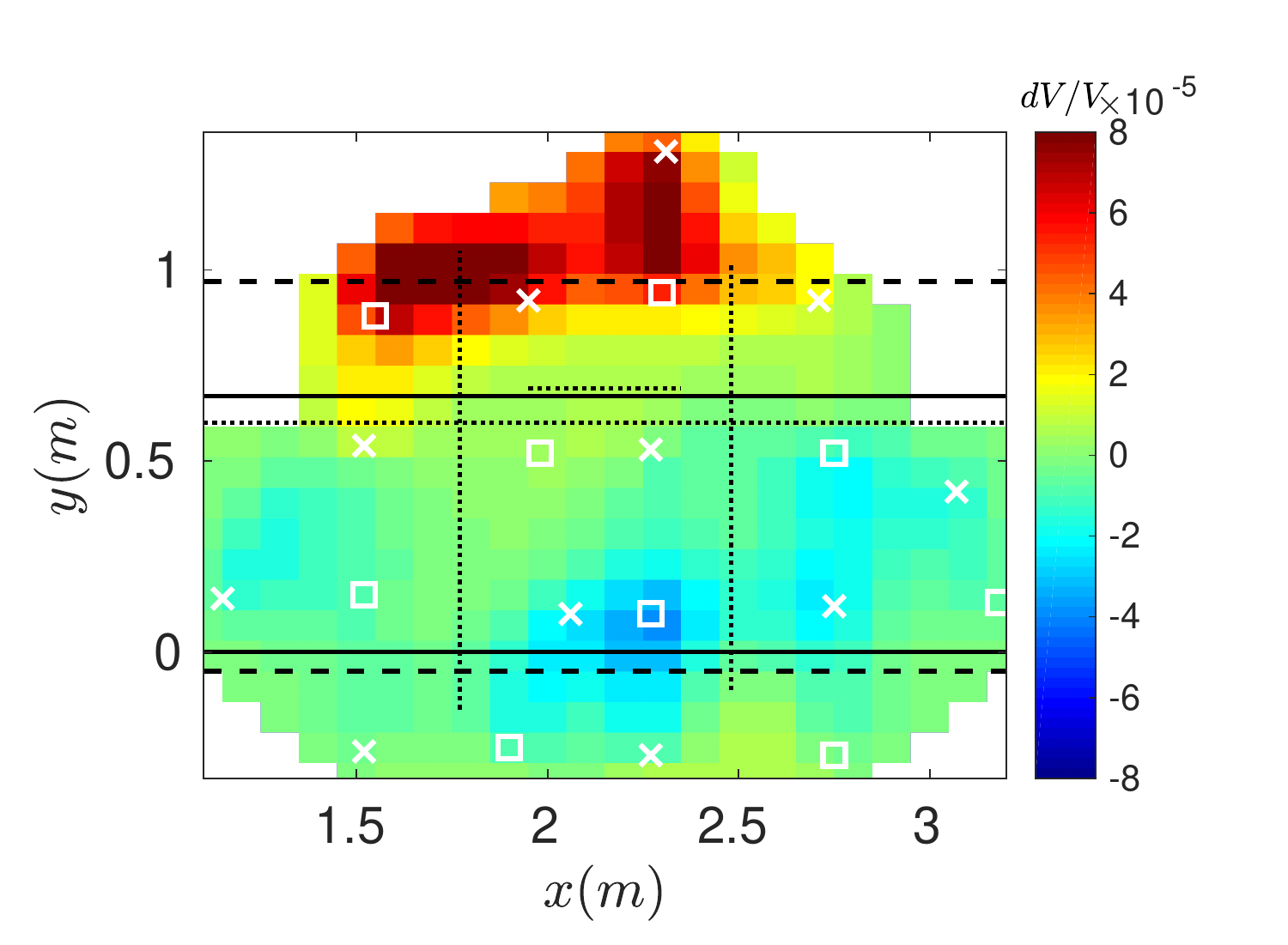}
   	\captionsetup{justification=centering}
   	\caption{}
\end{subfigure}~
\begin{subfigure}[p]{.33\textwidth}
    \centering
   	\includegraphics[width=\textwidth]{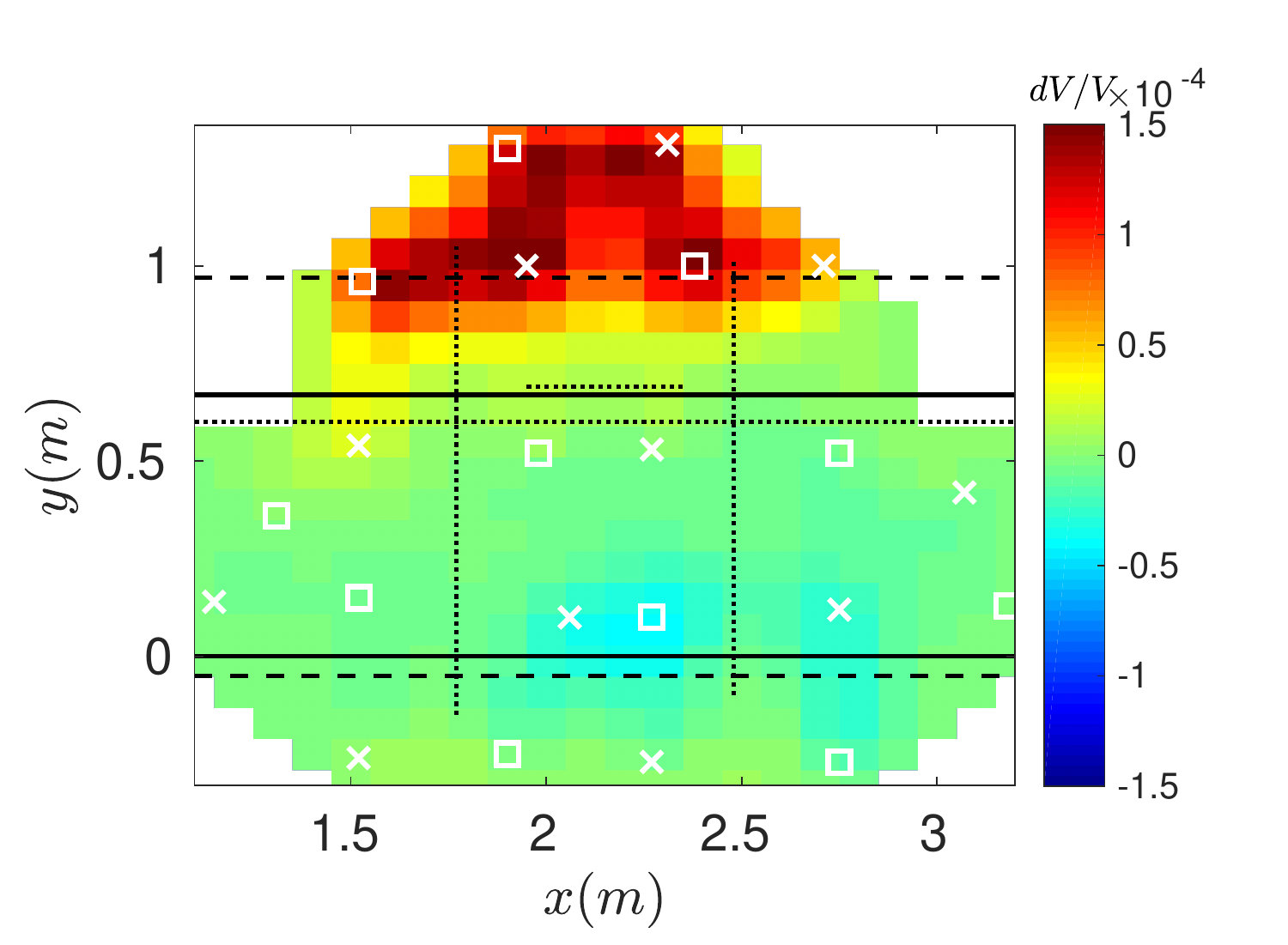}
   	\captionsetup{justification=centering}
   	\caption{}
\end{subfigure}
\caption{(a)-(c) average relative velocity change between consecutive measurements (one hour apart) at a depth of 5 cm during the second inflation stage in 2017, 2018 and 2019, respectively;
	(d)-(f) results during the first deflation stage.
	The white crosses and squares indicate positions of sources and receivers, respectively.
	Black lines are the same as those in Fig.~\ref{fig:config}.}
\label{fig:velocity}
\end{figure*}

\begin{figure*}[htb]
\centering
\begin{subfigure}[p]{.33\textwidth}
    \centering
   	\includegraphics[width=\textwidth]{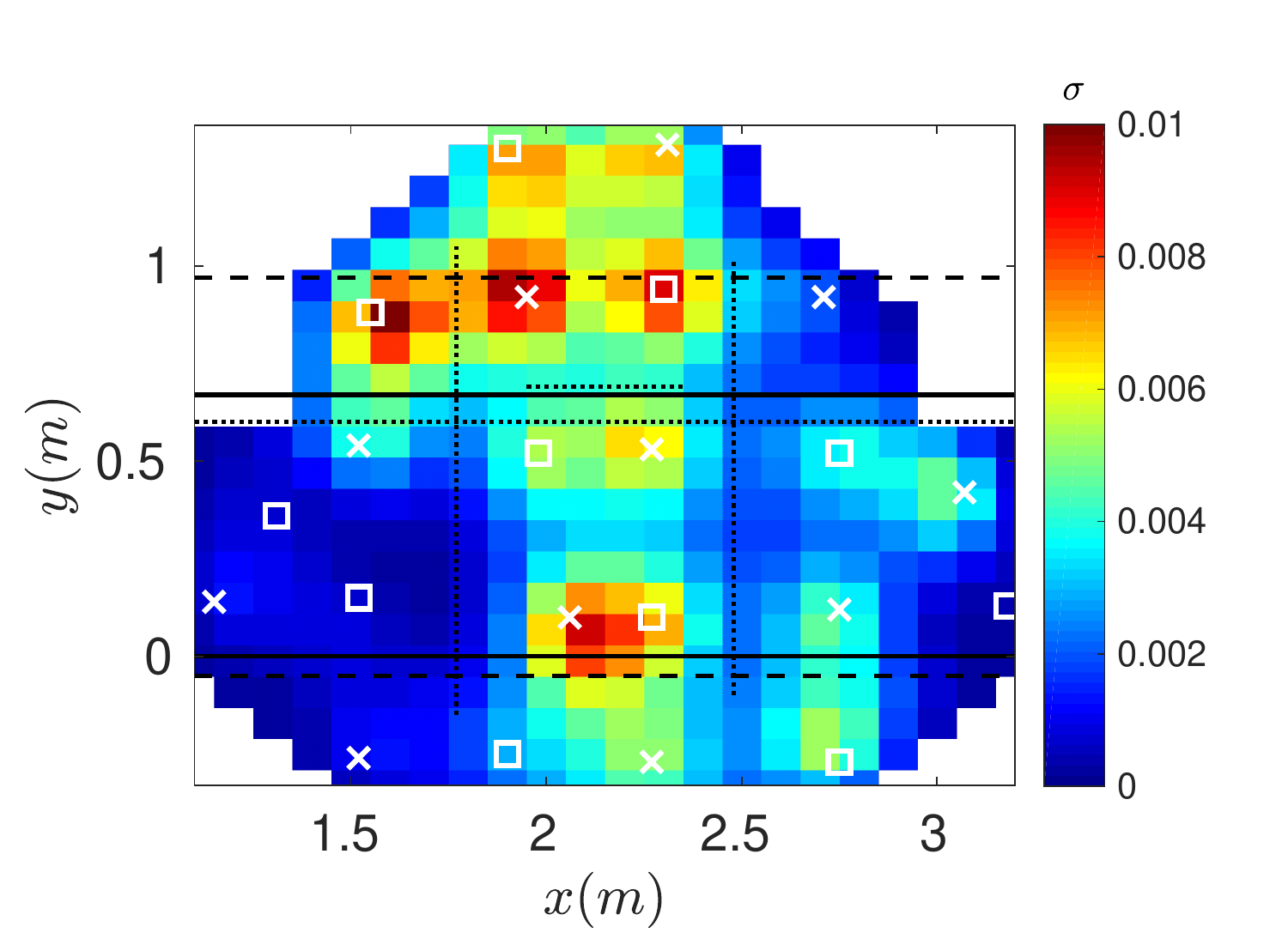}
   	\captionsetup{justification=centering}
   	\caption{}
\end{subfigure}~
\begin{subfigure}[p]{.33\textwidth}
    \centering
   	\includegraphics[width=\textwidth]{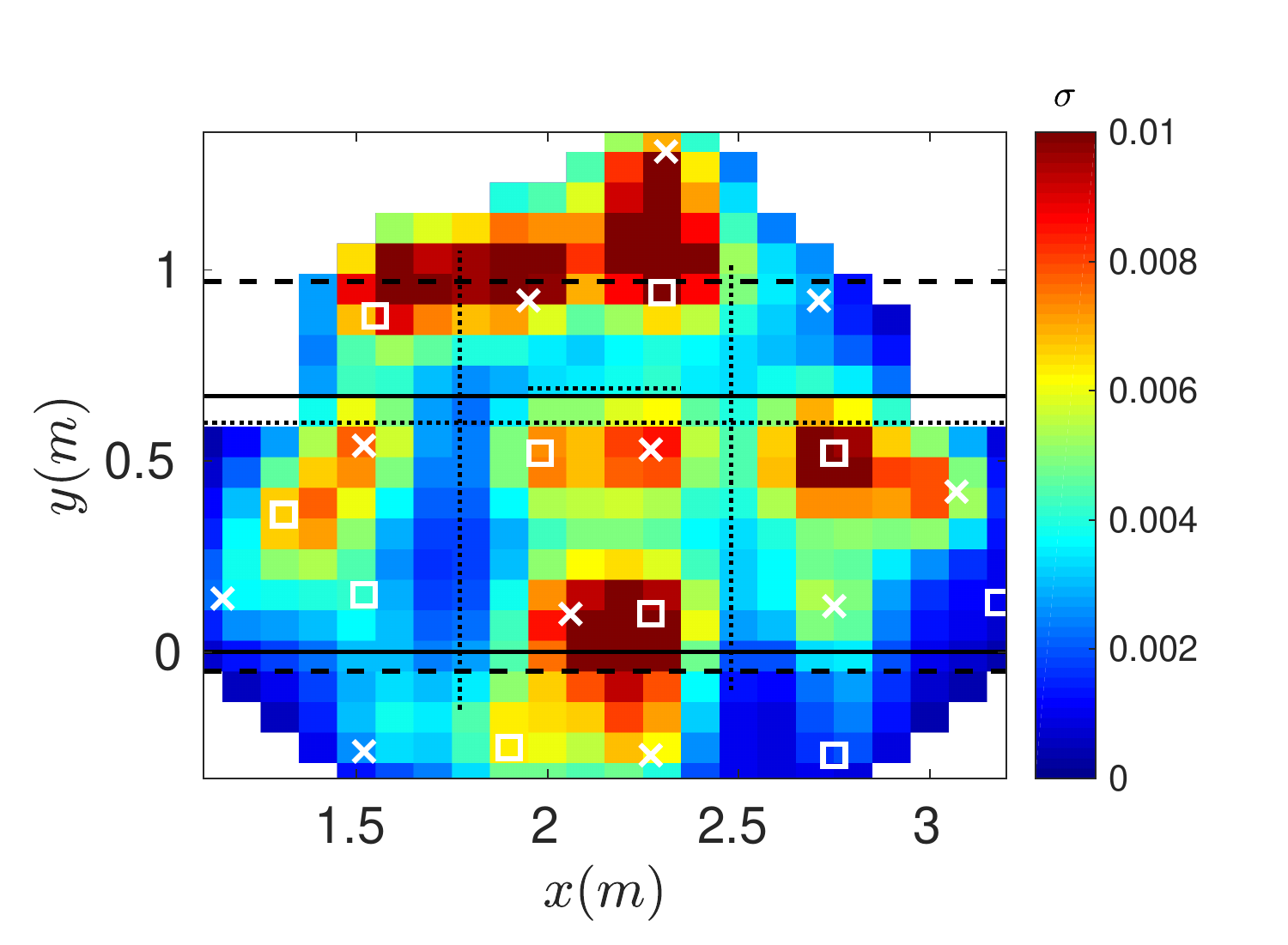}
   	\captionsetup{justification=centering}
   	\caption{}
\end{subfigure}~
\begin{subfigure}[p]{.33\textwidth}
    \centering
   	\includegraphics[width=\textwidth]{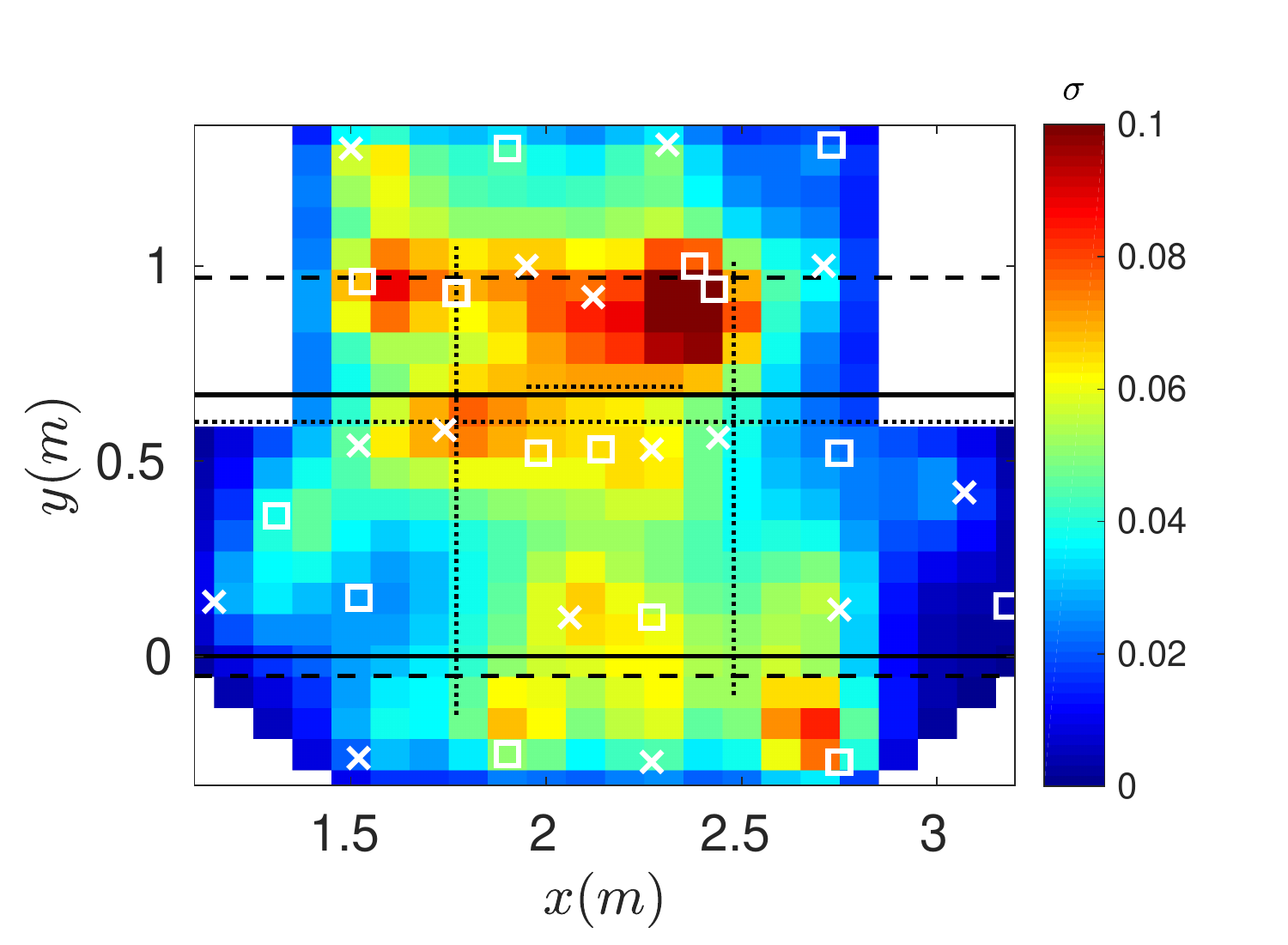}
   	\captionsetup{justification=centering}
   	\caption{}
\end{subfigure}
\caption{(a)-(c) average structural (scattering) change between two adjacent measurements at the depth of 5 cm
	during the second inflation stage in 2017, 2018 and 2019, respectively.
	The white crosses and squares indicate positions of sources and receivers, respectively.
	Black lines are the same as those in Fig.~\ref{fig:config}.}
\label{fig:cross_section}
\end{figure*}

The relation between the relative velocity change $\tfrac{dV}{V}(x)$ and $dV/V_a$ is given by
$$\frac{dV/V_a}{t} = \int \tfrac{dV}{V}(x)K(s,r,x,t) dx,$$
and the formula connecting the scattering change and DC has a similar form
$$\text{DC} = \int \tfrac{V\sigma(x)}{2}K(s,r,x,t) dx,$$
where $V$ is the average velocity of the ultrasound and $\sigma$ is the local scattering cross section density.
The reader may refer to \cite{planes2014decorrelation} for more details.
These two formula share a common sensitivity kernel $K(s,r,x,t)$, which is computed from 
the diffusive approximation of the coda wave. The sensitivity kernel reflects the probability that a wave emitted at $s$ at time $0$ and collected at $r$ at time $t$ has passed through the position $x$.
Here $t$ is in practice the center of the time window where we evaluate $dV/V_a$ and DC experimentally.
Refer to \cite{xue2019locating} for a detailed discussion about computing the sensitivity kernel and solving the integral equations.

We compute the velocity change between every two consecutive states and take the average value from all 
consecutive pairs in the second inflation stage (from 0.21 MPa to 0.41 MPa) and
the first deflation stage (from 0.40 Mpa to 0.21 Mpa) to generate Fig.~\ref{fig:velocity}. 
Each step has, in average, a change of + or - 0.02 MPa.
The decreasing (increasing, respectively) velocity is observed in the inflation stage in the top region
(in the lower middle region, respectively), and the opposite phenomenon occurs during the deflation stage.
The magnitude in 2019 is much larger than in 2017, which indicates again the cumulative damage of the concrete over years.
We only present velocity changes at a depth of 5 cm from the surface. 
We obtain similar results for other depth, except a decrease of magnitude for deeper layers.

Fig.~\ref{fig:cross_section} shows the average structural changes, or more precisely the scattering cross section density $\sigma(x,y)$ of the changes, between two consecutive states during the second inflation stage. The units ($m^2/m^3$) can be understood as a density of micro-cracks per volume unit. We observe an increase of scattering change from 2017 to 2019.
Our result shows that the most active region is along the top construction joint.
We also observe the developing of the crack near dotted lines from 2017 to 2019.

\section{Conclusion and discussion}
In this work, we use diffuse ultrasound to study the pressure and structural changes in the gusset area of 
a 1/3 replica of a nuclear power plant containment wall. Changes are induced by the increase and decrease of the inner pressure. We list the main results below:
\begin{enumerate}
\item We are able to image the ultrasonic relative velocity change with a magnitude as small as $10^{-4}$	thanks to the high sensitivity of diffuse ultrasound.
\item The stretching of diffuse ultrasound and reconstructed velocity change indicate that the concrete near the base is in compression 
	and in traction in the upper region in the inflation process, in agreement with mechanical models.
\item The increase of the slope of $dV/V_a$ versus internal pressure change from 2017 to 2019, in other words the sensitivity of  $dV/V_a$ to internal pressure, suggests that the concrete is progressively damaged and weakened. That is to say, the concrete deteriorates over time.
\item Although we do not detect significant pressure change in the base (bottom) region, 
	it could actually also suffer severe cumulative changes. The large DC value between the initial and final states 
	suggests that the concrete does not restore to its initial state during the test, i.e., the damage is permanent and irreversible, at least on the time scale of the test.
\item The reconstructed structural changes (scattering cross section) shows active regions during the inflation-deflation procedure that could correspond to cracking inside the concrete.
\end{enumerate}

\section{Acknowledgements}
This work was partially funded by the ANR ENDE (PIA 11-RSNR-0009 ) project, and by the IDEX Universit\'e Grenoble Alpes innovation Grant (CND-US).

\section*{Appendix A}
\setcounter{figure}{0}   
\renewcommand\thefigure{A\arabic{figure}}
In \cite{weaver2011precision}, authors proposed to evaluate the uncertainties of $dV/V_a$ due to the change in wave speed
through the computation of its root mean square, such that
\begin{equation}\label{eqn rms}
\text{rms}(dV/V_a) = \sqrt{\frac{6\sqrt{\frac{\pi}{2}}T}{\omega_c^2(t_2^3-t_1^3)}}\frac{\sqrt{1-\text{CC}^2}}{2\text{CC}}
\end{equation}

\begin{figure}[htb]
\centering
\includegraphics[width=.5\textwidth]{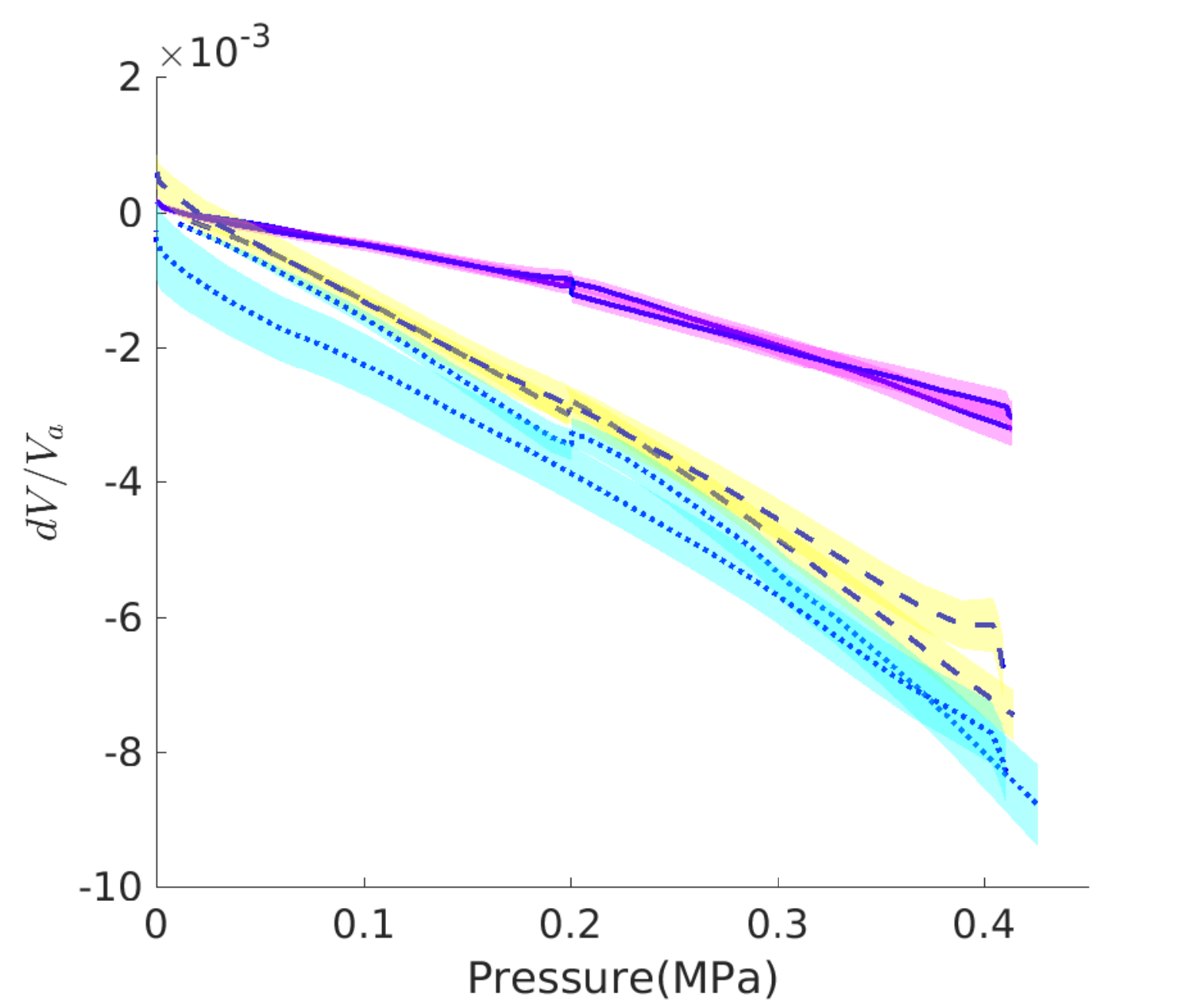}
\caption{Apparent velocity change calculated in 2017 (solid line), 2018 (dashed line) and 2019 (dotted line) from source-receiver pair at the upper region, and the corresponding uncertainties evaluated from Eq.~(\ref{eqn rms}) (colored region).}
\label{fig:dv_precision}
\end{figure}

For our experiment, the parameters are as follows:
\begin{eqnarray*}
\omega_c & = & 2\pi\times 1.5\times 10^5 \text{ rad/sec},\\
T        & = & \frac{\ln 10}{(2.2\times 10^5-0.8\times 10^5)\pi},\\
t_1      & = & 1.6\times 10^{-4} \text{ sec},\\
t_2      & = & 4.8\times 10^{-4} \text{ sec}.
\end{eqnarray*}
Application of this formula to data recorded in 2017, 2018 and 2019 at the upper source-receiver pair gives the uncertainties represented as colored region in Fig. \ref{fig:dv_precision}.

\section*{Appendix B}
Here we evaluate numerically the state of stress inside the concrete around the gusset.
We perform elastic linear numerical simulations using the ASTER code \cite{CodeAster}, without taking into account the possible role of cracks. For clarity, we only represent the difference of stress between the initial state (additional internal pressure  0 MPa) and the final state (internal pressure 0.43 MPa). 
Fig.~\ref{fig:7a} and \ref{fig:7b} represent a 2D vertical cut and a vertical profile of this tangential stress change. Positive values correspond to traction, where cracks are expected to develop and open, and where $dV/V_a$ is expected to be negative. Negative values of stress correspond to compression areas, where $dV/V_a$ are expected to be positive. Experimental results of $dV/V_a$ are perfectly in agreement with this mechanical simulations, demonstrating the ability of $dV/V_a$ obtained from ultrasonic coda waves to cartography the state of stress inside the structure.

\setcounter{figure}{0}   
\renewcommand\thefigure{B\arabic{figure}}
\begin{figure}[htb]
\centering
\includegraphics[width=.5\textwidth]{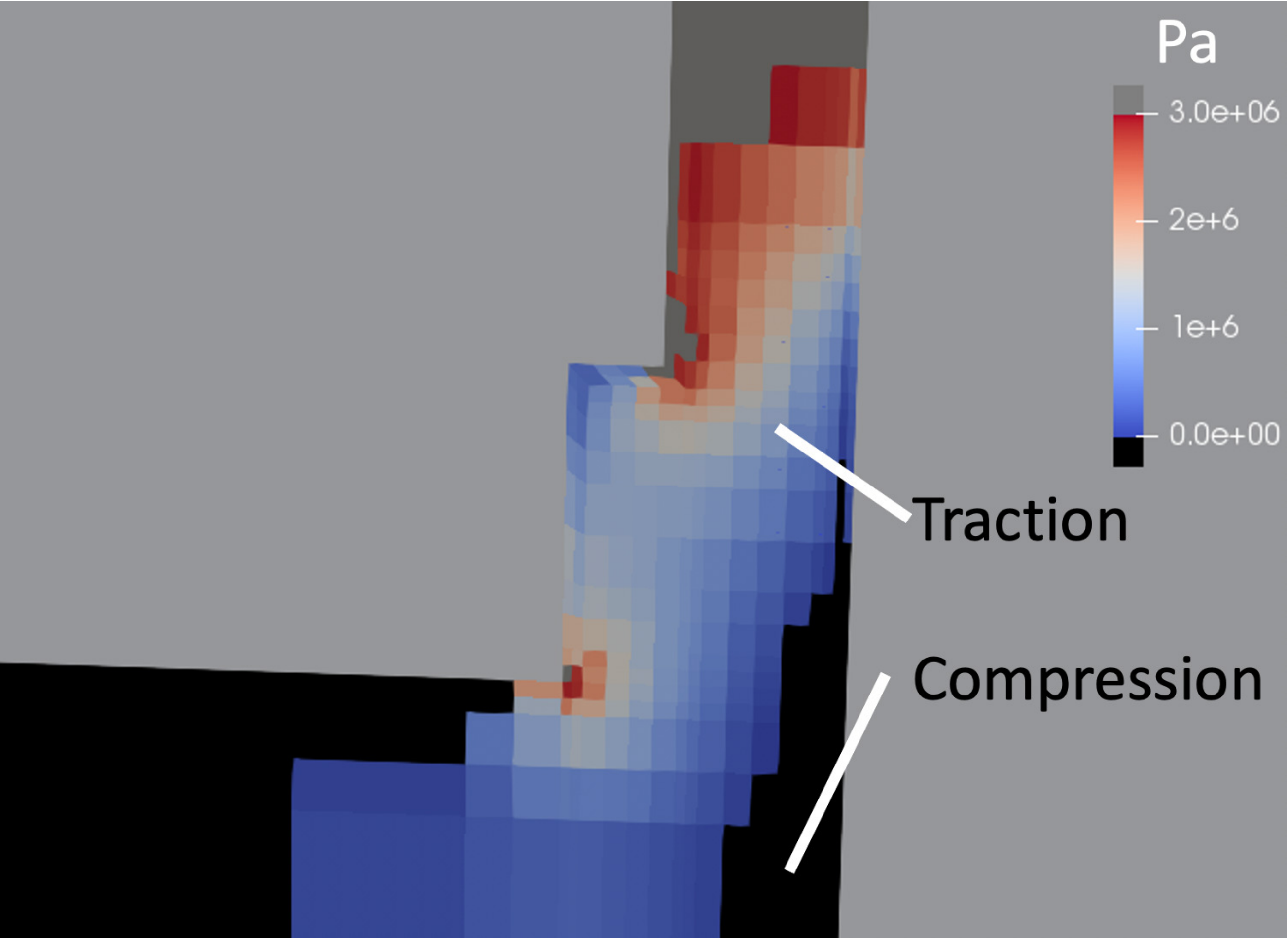}
\caption{Numerical simulation of stress around the gusset (vertical cut). We represent here the difference in tangential stress from the initial state (no internal pressure) to the plateau at 0.4~MPa. The colored areas indicate traction (form 0 to 0.3~MPa approximately). Black color indicates compression or neutral areas.}
\label{fig:7a}
\end{figure}

\begin{figure}[htb]
\centering
\includegraphics[width=.5\textwidth]{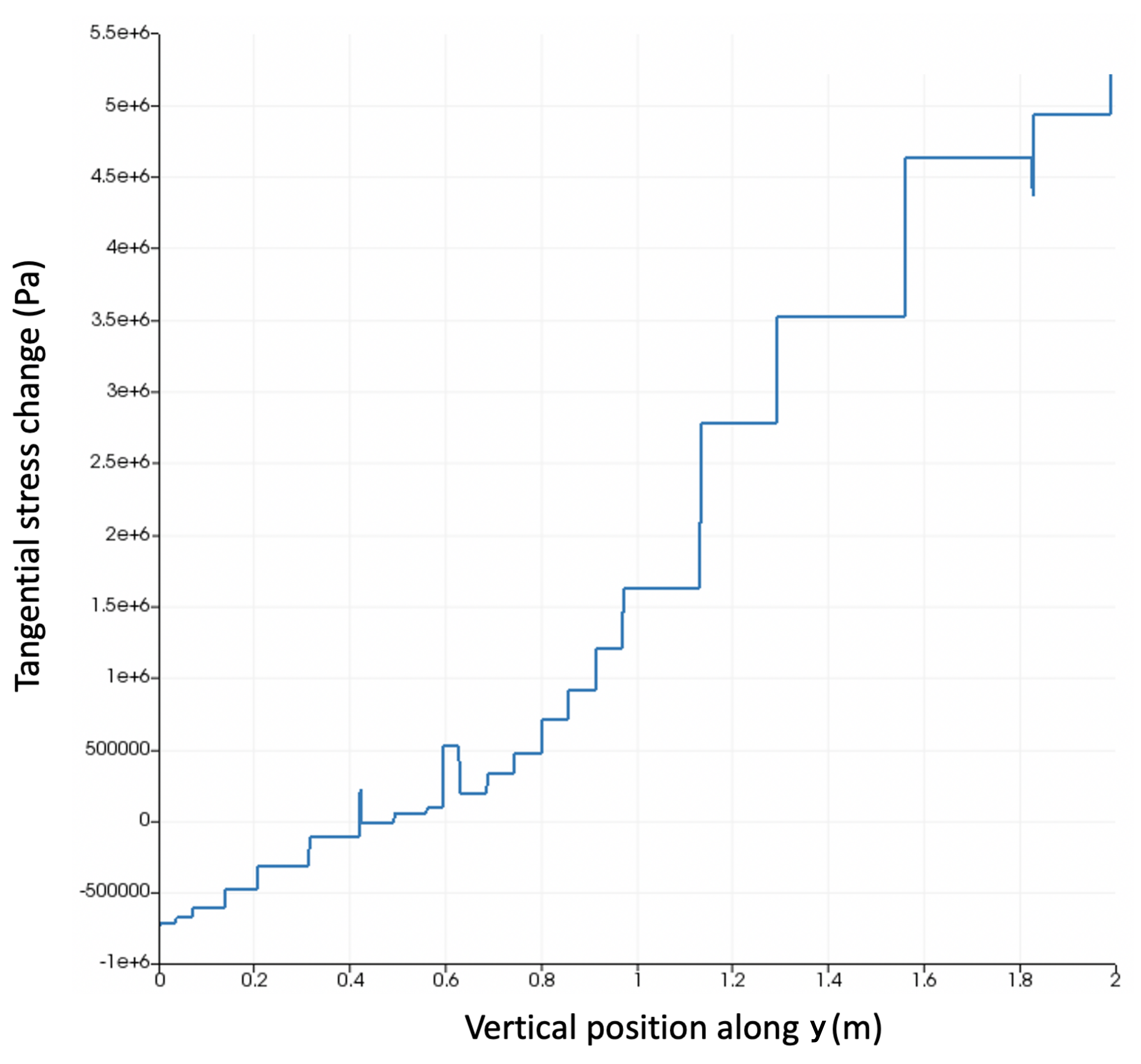}
\caption{Tangential stress difference between the initial (0~MPa) and final (0.43~MPa) state along a vertical line (y-axis) at a 5 cm depth inside the concrete (from the outside surface). Negative values hold for compression, positive values for traction.}
\label{fig:7b}
\end{figure}

\bibliographystyle{SageV}

\end{document}